# UNMANNED AERIAL VEHICLE FORENSIC INVESTIGATION PROCESS: DJI PHANTOM 3 DRONE AS A CASE STUDY[1]

Alan Roder
alan.roder@ucdconnect.ie

Kim-Kwang Raymond Choo
Department of Information Systems and Cyber Security, University of Texas at San Antonio, TX 78258, USA
raymond.choo@fulbrightmail.org

Nhien-An Le-Khac
School of Computer Science, University College Dublin, Ireland
an.lekhac@ucd.ie

**ABSTRACT**

Drones (also known as Unmanned Aerial Vehicles – UAVs) is a potential source of evidence in a digital investigation, partly due to their increasing popularity in our society. However, existing UAV/drone forensics generally rely on conventional digital forensic investigation guidelines such as those of ACPO and NIST, which may not be entirely fit-for-purpose. In this paper, we identify the challenges associated with UAV/drone forensics. We then explore and evaluate existing forensic guidelines, in terms of their effectiveness for UAV/drone forensic investigations. Next, we present our set of guidelines for UAV/drone investigations. Finally, we demonstrate how the proposed guidelines can be used to guide a drone forensic investigation using the DJI Phantom 3 drone as a case study.

## 1. INTRODUCTION

Drones, also referred to as Unmanned Aerial Vehicles (UAVs) in the literature, can be loosely defined as an aircraft piloted by remote control or an on-board computer. There are a wide range of UAVs, in terms of capabilities and prices. Such UAVs are also designed for use in different environments, such as security, disaster response (e.g. rescue missions), mapping and adversarial settings (e.g. battlefields).

UAVs can be considered as part of the broader Unmanned Aerial System (UAS), which encompasses UAV, Ground Control Station (GCS) and Controller. These parts are necessary to successfully, remotely and accurately control a UAV.

In recent years, UAVs have been increasingly popular among consumers and the research community. For example, the global market revenue for drones is expected to surpass $11.2 billion by the year 2020, according to a report from Gartner [1]. With so many drones purchased for home and personal use, the potential for drones to be involved in a digital (forensic) investigation will undoubtedly increase. For example, it was posited that vulnerabilities in driverless vehicles may be exploited by criminals, particularly terrorists, to

---

[1] Certain commercial entities, equipment, or materials may be identified in this paper in order to describe an experimental procedure or concept adequately. Such identification is not intended to imply recommendation or endorsement by the authors or their institutions, nor is it intended to imply that the entities, materials, or equipment are necessarily the best available for the purpose.

facilitate criminal or terrorist attacks in the physical world [2]. The same can be said for drones [7].

UAV forensics is relatively less studied, in comparison to other popular consumer devices and technologies such as mobile devices (e.g. Android, iOS, and Windows Phones), cloud computing, edge computing and fog computing [25].

In 2015, Kovar [3] highlighted the essential elements akin to UAV forensics, and detailed the process of obtaining data from the popular DJI Phantom 2. A year later in 2016, Kovar, Dominguez and Murphy [4] extended the prior work in [3] to include a forensic examination of DJI Phantom 3. Along a similar line, Horsman [5] conducted a forensic investigation of Parrot Bebop UAV, and Clark et al. [8] presented their findings of a Phantom 3 UAV forensic examination.

On the other hand, more than a decade ago in 2007, the Association of Chief Police Officers (ACPO) published 'The ACPO principles for obtaining digital evidence' [10]. In the same year, the National Institute for Science and Technology (NIST) published the 'Guidelines on Mobile Device Forensics' [11]. Existing UAV forensic approaches are generally based on ACPO and NIST guidelines (or their variations). This is not surprising as there is no published guideline designed for UAV forensics.

Hence, in this research, we review existing (UAV) forensic literature and potential data storage locations. In our review, we highlight the limitations in existing guidelines, and the need for a guideline dedicated to UAV forensics. Thus, we propose in this paper a forensic process focused on UAV investigations. This process is designed to guide the investigation process when examining UAVs.

We then evaluate the proposed process using a drone as a case study, and specifically a DJI Phantom 3 drone.

The rest of this paper is structured as follows. In the next section, we discuss UAV forensic challenges and briefly review existing forensic guidelines in the context of UAV forensics. We present our UAV forensic process in Section 3, and the case study in Section 4. We conclude and discuss future work in Section 5.

## 2. UAV FORENSIC CHALLENGES

UAV forensic and security examinations have been undertaken by UAV enthusiasts and the fan communities. For example, a number of them have created their own (often freely available) software, which can interpret the data files stored on the UAVs. One such example is DatCon, a tool designed to interpret .DAT files specifically from DJI UAVs [9]. While these tools are a valuable pool of knowledge, such tools are unlikely to have been validated according to forensic requirements. In other words, these tools are unlikely to be forensically sound and artefacts obtained from using such tools may be inadmissible in a court of law. Thus, there is a need for forensic validation work to be undertaken by the digital forensic community.

In addition to the diversity / variation in UAV products, it is understandable that the existing forensic examination guidelines may not be appropriate or sufficient. For instance, the ACPO principles for obtaining digital evidence [10] and NIST Guidelines for mobile phone forensics [11] were both published in 2007, and these guidelines may not have kept pace with technological advances.

In the context of UAVs or UAS, for example, data can be stored in several locations, such as the UAV, GCS, network routers, and so on. Storage locations can also be overt or covert, and one also needs to note that in some instances, there are in-built persistent storage media such as Micro SD cards [12]. There is also the likelihood of the recovery of artefacts from flash storage, which typically requires some form of direct connection [24]. We would also have to take into consideration the likelihood that a UAV used in a criminal activity has been modified to either hinder forensic investigation or enhance certain features such as increased load carrying capacity (e.g. in drug smuggling activities across borders, or act as an improvised explosive device).

As previously discussed, there are a number of existing digital forensic guidelines. When the ACPO principles [10] were created, it was an attempt to standardize what was then a relatively new field of forensic study. The four ACPO principles were generalized so that they are technologically neutral. However, it is important to note the key concept underpinning these principles is to ensure the integrity of the original data. This clearly applies to UAV and any forms of digital forensics.

There are also similarities between UAV and mobile device forensics [26]. For example, similar to a mobile device, a modern or advanced GCS is likely to have Wi-Fi, Bluetooth or Internet connection. Therefore, there is a possibility that the device could be remotely wiped or modified. UAV forensics can also involve conventional storage media forensics [24] (e.g. memory cards are copied) and live forensics (e.g. real-time access to a live UAV to view data stored on flash memory). Since most UAVs do not have a graphical user interface (GUI) or inbuilt interface, there is a real-risk that data may have been changed without the knowledge of the forensic examiner / investigator. Thus, consideration must be given at this level of examination, and while deciding the order of investigation one needs to minimize any potential for data modification. Since checking of UAV flash memory requires a live interaction, it is unlikely that any two examinations will achieve the same result.

Whilst existing literature is useful to guide a general forensic investigation of a UAV, having a UAV focused / specific forensic process could be more useful to forensic examiners / investigators (e.g. to maintain consistency across cases).

## 4. PROPOSED UAV FORENSIC INVESTIGATION PROCESS

In this section, we first determine if there are any differences between digital storage locations, when compared to traditional computer/mobile forensics. Next, we propose a new forensic investigation process for UAV.

### 4.1 UAV data storage location

In many ways, the storage locations for UAVs share similarities with mobile devices. UAV storage locations vary, but the medium used to store data is primarily either a Micro SD card or flash memory. This seems to be an over simplification given the constant evolution and advances in related technologies. For example, older mobile devices relied upon flash storage for operating system (OS) storage and Micro SD cards for additional storage. Since 2015, most mobile devices use flash storage. Given the demand for UAVs to become more efficient, it seems likely that they will follow a similar technological trend to mobile devices.

At the time of this research, popular commercial UAVs provide OS via flash storage or Micro SD card, with a separate Micro SD card for video footage. This flexibility allows the base UAV cost to remain low, whilst allowing upgrades to storage at the owner's expense. Since there is demand for UAVs to remain in flight for longer periods of time, and to provide increased 4K support for video capture, the likelihood that flash storage will become an option (similar to Apples graded internal storage pricing) becomes more likely.

A more significant variance between UAVs and mobile devices is the inherent adaptability and modular nature of UAVs. UAVs can store data in different locations such as the UAV, GCS, and other mobile devices used to connect/pilot the UAVs. Flight log data is often stored in a single location; however, media files are often found in multiple locations, usually in different resolutions.

Investigation on the data obtained from mobile devices, laptops and personal computers usually incorporate elements of registration information, such as email addresses, usernames and payment plans. Since UAVs traditionally do not require registration or payment plans [13], this further dilutes the association between the device and the operator.

### 4.2 Proposed process

Now we describe our UAV forensic investigation process step-by-step, using a case study for illustration. In our process, there are three main stages, namely: preparation, examination and analysis/report. The first stage includes Steps 1 to 6. Steps 7 to 17 are part of the second stage, and the final stage includes Steps 18 to 20.

*Step 1 - Identify and determine the chain of command*

Relevant questions to consider are as follows:

1. How is the exhibit seized? For example, has a tamperproof evidential container / bag been used, and have photographs been taken of the exhibit?

2. Has consideration been given to electronically isolating the exhibit (e.g. the use of Faraday box / cage)?

3. Does the container state the exhibit reference?

4. Does the container name the seizing officer or exhibitor?

5. Does the container have a unique reference number?

6. Does the container state when and where the exhibit was seized?

7. Does the container have sufficient space to sign your name?

The above questions are not an exhaustive list of considerations and should be adapted based on the situation, and the guidelines and rules of the investigation authority. As with any forensic examination, if the credibility of the exhibit cannot be maintained, its evidential usefulness will become limited. If the exhibit continuity is weak, then it creates an element of doubt in the admissibility of the evidence and a potential for the defence team to discredit part or all of the evidence obtained.

It is often the case that the UAV will be seized first (e.g. due to device failure or pilot error). Should data relevant to the case be obtained, there may be a lag between when the UAV was seized and when the warrant or arrest was executed. Should a GCS be found during the search process, it will need to be examined to determine if it is linked to the UAV.

Due to the inherent remote access associated with UAVs, consideration must be given to network isolation. It may be safer, cheaper and more practical to switch off the UAV at the point of seizure. However, since there does not appear to be a standard OS across the wide range of UAVs, consideration must be given to how data is stored and what effect this will have.

*Step 2 - Have conventional forensic practices (e.g. DNA, fingerprints, and ballistic) already been implemented?*

Digital evidence can also be supported by traditional evidence such as witness statements. For example, fingerprints and other DNA materials found on a UAV can also be used as supporting evidence in the investigation.

*Step 3 – Identify the role of the device in conducting the offence (Offence analysis)*

This step includes two important tasks, namely: (i) Review the case investigation notes to determine how and why this device was used during the commission of the offence; and (ii) Identify what the offence was and how it is alleged that the UAV was used.

In other words, we need to recover artefacts to support the elements of proof, and thus focus our forensic investigation accordingly. For example, if it was alleged that a drone was used during a voyeurism offence, then the drone's video footage, etc may be more useful evidence than flight logs.

*Step 4 - Photographs*

During any digital examination, photographs should be taken. These photographs may help to prove beyond reasonable doubt that the exhibit was in the condition described during the notes. Device images should be taken which present the following:

- Exhibit within the tamperproof container.

- The tamperproof container, including exhibit reference, unique seal number, etc.

- Exhibit out of the tamperproof container.

- Exhibit from all possible angles.

- Any markings or serial numbers.

- Any obvious modification.

- Any damage(s).

- BIOS, if possible (and this can be performed later during examination, when data storage media has been removed).

- Load carrying mechanism, if applicable.

- Defensive / offensive capability, if applicable.

During the examination, the device BIOS data may be obtained. When images are taken, these should also contain a digital radio clock with the current date and time. Photographs should be taken to accurately portray any load carrying mechanism (where one can be identified). Also if defensive or offensive capability has been identified, then consideration should be given to the safety of the examiner and sufficient precautions be made to prevent any injury. The list of photographs which should be taken is not exhaustive. In principle, photographs should be taken of any relevant aspect of the exhibit that may prove evidential, either in supporting or refuting the supporting evidential material or assumptions.

*Step 5 - Identify the make and model*

At this step, identification should be via a visual inspection, taking into account markings, designs and patterns, and cross referencing. Such identification can be facilitated through experience and open source researching. Identification can help the investigation in a number of different ways. If the device has a high value, then consideration should be given to whether or not it is a stolen device.

Using local law enforcement resources, it may be possible to create a short list of recent thefts and burglaries where a UAV was stolen. Should a suspect already have been identified for the theft, then this may present investigative avenues to help identify the UAV operator.

Whilst this step is not entirely unique to UAV forensics, it is likely to be far more common. Computers and mobile devices are usually seized from an address or individual, however UAVs are more likely to be seized when the operator is not nearby or at a crime scene, as such attribution becomes more difficult.

*Step 6 - Open source investigation to identify device characteristics, potential data storage locations, and available forensic / non-forensic tools*

*Device characteristics:* Identify if a device is genuine or a counterfeit by identifying the markings, light locations and any other significant feature(s) and comparing such information against the specifications of the product as listed on the manufacture website.

*Potential data storage locations:* Whilst some memory card locations will be clearly marked and easily accessible, some may not. Some devices have removable storage, whilst others may have inbuilt flash memory. Understanding the potential locations for data storage will allow one to plan the forensic examination and reduce the possibility of missing evidence.

*Available forensic / non-forensic tools:* Many of the analysis tools which will be used, will likely have been created by drone enthusiasts. There are currently only a few UAV-specific commercial forensic tools available (e.g. Cellebrite and MSAB); however, their portfolio of models catered for is limited.

This step is not entirely unique to UAV forensics, but given the limited forensic literature available in this field, it is a key feature.

There is no standard location or format for UAV flight data, and research is necessary to prevent missing evidence or misinterpreting extracted data.

*Step 7 - Identify capabilities (Video/Audio recording, carrying capacity and technique)*

The following two steps are arguably the first stages which are entirely unique to UAV forensics, when compared to other forms of digital forensics. The reason being that, for example, most commercially available UAVs are not designed to carry payloads and release them. Commercially sold UAVs are 'currently' not designed to hold a firearm or offensive weapon.

Since UAV investigations will likely be related to a criminal offence, it becomes more crucial to determine how the UAV was used and what (if anything) was adapted to allow the UAV to carry out the offence.

It is unlikely that one would detail and highlight modifications to a desktop computer or a mobile device, since neither are historically used beyond what they were designed to do.

In this step, investigators aim to answer the following questions:

- Does the device have a video capture facility (see Figure 1a)?

- Does the device have an audio capture facility?

- Does the device have a load carrying capacity (see Figure 1b)?

- Does the device have an offensive capability (see Figure 1c)?

- Does the device have a defensive capability?

It is recommended that one conducts a visual examination of the device and takes note of each of its capabilities, as well as taking photographs where appropriate. Where offensive or counter-offensive capabilities are noted, consideration must be given to minimize health and safety risks to the examiner, and appropriate safeguards should be put in place. A criminal investigation can change direction, based on new information uncovered.

*Step 8 - Identify potential modifications.*

The standard drone specifications (depending on the drone) are sufficient for the task they were designed to complete. The use of UAVs in the commission of some criminal offences may require modifications to the UAVs, as previously discussed.

Thus, identifying such modifications will help support an investigation to either confirm or refute the alleged use during the offence. An example could be the sending of items into a restricted area (e.g. prison). Most standard drones do not have a load carrying mechanism. Due to flight time restrictions, the drone may have a non- standard battery (to increase flight time). The drone may also have non-standard motors to reduce noise levels.

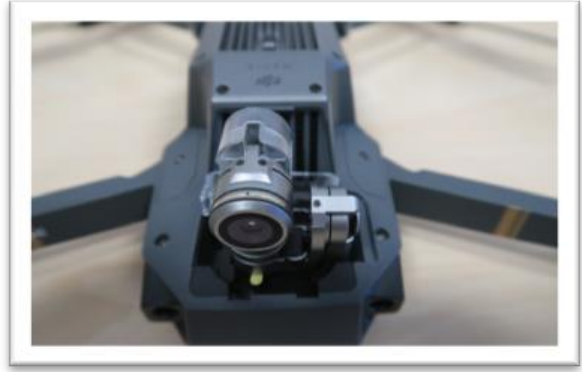

(a) UAV with a camera (HD camera fitted to a DJI Mavic Pro).

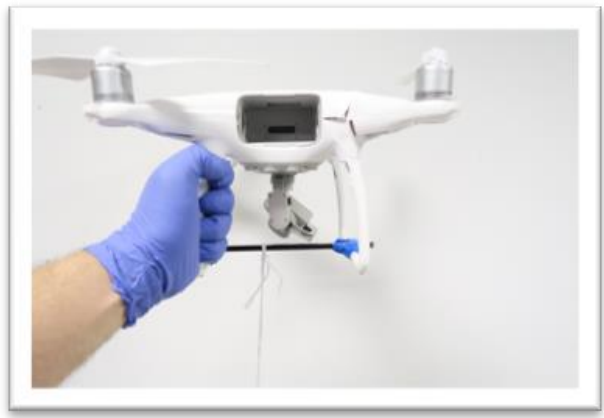

(b) UAV load carrying (Image shows a DJI Phantom, where a rod of plastic has been taped spanning the legs. A string can be seen hanging down, which would have held the payload).

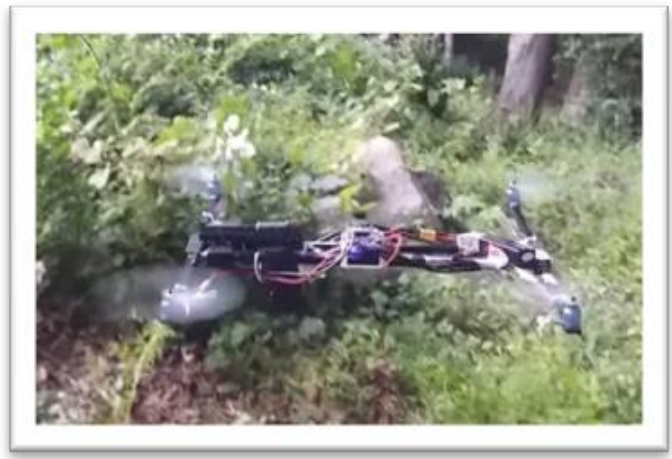

(c) UAV with an offensive capability (Image shows a custom built UAV with a 9mm pistol attached to the frame, which was taken from a video showing the pistol firing whilst the UAV was airborne).

Figure 1. Identify capabilities

Items of interest include non-standard battery, non-standard motors, non-standard propellers, non-standard camera, and load carrying device.

Identifying the standard characteristics of a UAV can prove tricky, since not all manufactures list all of the parts present. Consideration should then be given to either contacting the manufacturer directly and/or expanding the sources of information to include enthusiast forums and similar websites.

*Step 9 - Identify data storage locations.*

Relevant data storage locations in a UAV include removable memory card (SD, Micro SD, etc.), fixed memory card, flash memory (NAND, NOR, etc.), and SIM card.

Drone data storage locations can vary considerably and, in some cases, data can also spread over multiple locations. Some drones will capture media and store the original version on the drone, whilst also streaming a reduced quality version onto a storage device (e.g. mobile device or the cloud). Some drones will have visible slots, which are designed to allow easy access and swapping of portable storage devices (memory cards). Often these will be the default storage location for media. Some drone models will have hidden and potentially sealed portable storage devices (memory cards). Often these will be the default locations for system information and potentially flight logs.

Below are two drone models with model-specific storage capabilities:

- The DJI Phantom 4 has two removable data storage locations on the drone. The first contains media data, whilst the second contains flight log data (including ancillary data such as motor speeds) [14].

- The Yuneec Typhon H has one data storage location on the drone, which contains media data [15]. The flight log data is stored on the dedicated GCS [16].

*Step 10 - Identify ports*

There are a variety of different methods that can be utilized to enable interaction with a drone, and external ports appear to be the most common method used by manufacturers. External ports such as USB (2.0/3.0), USB-C, Micro USB and Lighting can potentially allow access to a drone's data storage, where storage is considered to be either flash or fixed.

Consideration should be given to conducting this type of examination, as it will likely involve powering on the exhibit. Any examination of this type will require an understanding of the drone systems, as data will likely change.

Should evidence be obtained, the examiner will need to be able to explain what data change during the examination, and why the evidence obtain during the examination can be relied upon.

*Step 11 - Extract removable data storage mediums*

In this step, we recommend the use of non-destructive methods. Consideration at this stage should be to extract only data sources that do not require destructive methods (e.g. chip-off). Destructive methods should only be considered when all other methods fail. As with any forensic examination, notes must be made to identify where removable storage devices were taken from. These storage devices will then need to be sub-exhibited in accordance with the naming conventions stated by the examiners force.

*Step 12 - Preserve evidence – Clone / forensic copy of storage medium*

This is a common practice in many digital forensic examinations, as such the process will not be explained in this document. It should be noted that cloning a removable storage medium may be beneficial when attempting to access data which may otherwise be unobtainable.

By way of an example, the DJI Phantom 4 flight logs are stored in .DAT files. These are normally classed as 'Generic data files', but unlike most file types they do not have an associated software to read them. When a new flight log is opened, it also has the secondary effect of closing the previous .DAT file. The last flight log is not viewable until the device is turned on.

By cloning the removable storage device, the examiner is then able to replace the memory card with the cloned memory card, power on the device; thereby, closing the final .DAT file, and ultimately re-examining the memory card which now has the last recorded flight data (last recorded prior to seizure). Original data has not been changed, but new data has now become viewable.

*Step 13 - Traditional interrogation of storage medium - use certified forensic tools*

This is a common practice in many digital forensic examinations. It should be noted that traditional forensic tools may successfully extract media files; however, flight logs may show as 'unreadable'. UAV manufacturers may store data in different formats, and currently there is no standardization. Should any data be identified, consideration must be given to checking the data though another tool and confirming that it has been interpreted correctly.

*Step 14 - Extended interrogation of storage medium*

This step is somewhat unique to UAV forensics. Typical digital forensic analysis is normally conducted using commercial forensic tool, which will usually have a proven record for accuracy. Any examination using non-validated tools is considered a risk. However, until commercial forensic tools for all UAVs are available, we may have little choice but to rely on open source tools to extract data of forensic interest.

As previously discussed, the capabilities of such open source tools can vary significantly. In some cases, extracted data can provide significant information, whilst others may only provide limited data. Examples of such tools include DatCon (Primarily DJI) [9], DJIFix (carves images and videos through the command line) [17], st2dash [18], and DroneLogbook [19]. There are both advantages and disadvantages in the use of such tools:

Advantages

- If the UAV stores data in a format that is supported by existing tools, then the open source tool can often interpret data that may not be understood using existing tools.

- Tool is also freely available, although one should note the software fair usage restrictions.

- Most makes and models are supported (with varying success).

Disadvantages

- The data obtained is unverified, incomplete or corrupted.

- Tool updates are sporadic or non-existent.

- Previously available tool may be removed without warning.

- Increased risk of obtaining malware.

When considering potentially non-validated open source tools, validation of results will prove necessary [23].

*Step 15 - Interrogation of the UAV / drone - Potentially using a clone of any storage medium identified*

In certain circumstances, it may not be possible to remove storage devices, such as embedded multimedia card (eMMC) storage. Prior to conducting destructive examination techniques (chip-off, etc.), consideration should be given to performing live examination of the device.

The most common connection is via direct cable and this will likely be the case for the immediate future. For example, advances in mobile technology and digital forensic tools may result in other remote ways of obtaining evidence from UAVs. One should be open and research for potential connection methods appropriate to the UAV in question.

There may also be product specific software available that supports device examination; however, consideration should always be given to the validity and forensic soundness of the tool.

*Step 16 - Interrogation of peripheral devices: flight controller, mobile device, etc.*

Streaming data techniques and cloud storage mean that data may no longer be stored on the physical device (beyond system and function files). Thus, consideration should also be given to ancillary devices that may be used to control the device and/or store data.

Most drones require a GCS in some form, which can take the form of a dedicated GCS (handset), mobile device (phone, tablet etc.), laptop or potentially a computer that could input flight paths without the need for later remote access, which could perform autonomous flight.

This step becomes more specific to UAV forensics, since it could be considered that UAVs form part of the overall exhibit, which is the Unmanned Aerial System (UAS). We do not consider a mobile device to have independent components, nor a computer. However, the media consistently refer to a UAV (or Drone), but never the UAS, as such this stage is significant.

The examination of such devices could be conducted using known forensic tools (Cellebrite, MSAB, etc), but this process may only be limited to mobile phones and tablets which are usually supported. Equally computers could be examined using tools such as EnCase or X-Ways.

The larger issue arises if the forensic tool does not understand or cannot interpret the file(s) holding the required data. Whilst this guide focuses on UAV examination, steps 6, 10, 11, 12 and 14 would also support GCS examination.

When considering GCS in the form of mobile phones and tablets, we also have to consider applications, since this would be the likely platform used to interact with the UAV. Whilst further work around peripheral devices would clearly support and enhance this guide, at this stage there are too many variables (due to a lack of standardization) to include.

*Step 17 - Extract removable data storage mediums (Destructive)*

Destructive extraction methods such as chip-off should be considered a final resort for obtaining data from a digital storage device. Should this extraction method fails, the likelihood of obtaining useful evidence will be significantly reduced unless another method becomes available at a later date.

From the authors' experience, it would appear that the preferred storage medium for UAVs is a micro SD card. However, if the UAV follows the same technological curve as mobile phones, this could be replaced with an eMMC. At that time destructive methods may become more likely unless the UAV. drones being examined are supported by tools available at the time of examination.

*Step 18–Initial review of extracted data*

Analysis largely depends on the offence under investigation and the elements of proof required. For example, image and video metadata may hold file creation times and dates, along with GPS data. Interestingly, due to the often hidden location of data storage locations, suspects may fail to remove previous data prior to committing offences. For example, our personal experiences have shown that often the first images stored on the device are images of the suspect playing with the device and learning how the recording function works.

Flight data can also vary dramatically with some recording little or no data, whilst others will record GPS position (including altitude), individual motor speed, pitch and yew and a whole host of details and photographs. For example, flight logs generated through examinations of DJI UAVs (using the DatCon analysis software) showed significant detail, including motor speeds and battery usage. Conversely, flight logs generated through examinations of the DJI GO application on a tablet (using Cellebrite software) showed much less data.

*Step 19 – Interpreting and translating of data - Into a human readable and evidential format*

Digital examinations can, and often do, produce a significant amount of data. There are three main aspects to this step, which can be broadly categorized as data sifting, data confirmation and data translating:

- Data sifting is the process of reducing the data obtained through examination, to only case relevant data.

- Data confirmation is the process of verifying the obtained data and confirming its accuracy.

- Data translation is the process of changing often complex data sets into a human readable format.

*Step 20 – Report/Statement*

A sound report can have a significant influence on the likelihood of a conviction and/or sentencing. Specifically, a well-written report should focus on the facts of an examination and its conclusion should be an impartial assessment of the data obtained through an examination. For example, where flight log data has been extracted and extrapolated, consideration should be given to providing a visual representation of the location. Whilst many options are available, often a simple mapping software will suffice. Consideration should also be given in relation to unnecessary information.

It should also be noted that while the guidelines flow sequentially, each part can, or possibly, should be considered as independent, and can be conducted at any point in an examination as required.

## 5. A CASE STUDY

In this section, we demonstrate how our forensic process presented in Section 4 can be used to guide the forensic investigation of a DJI Phantom 3 UAV (see Figure 2).

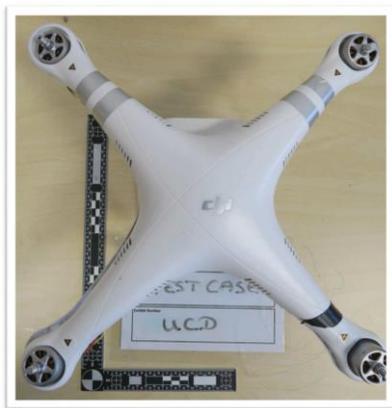

Figure 2. DJI Phantom 3

Since the experiment was designed to evaluate the process of obtaining data, rather than analyzing the data extracted, the examination notes do not contain personal data relating to the UAV (or where data is represented, it is edited). No report will be generated which contain flight logs.

As this is a used case exhibit, we will skip the requirement for a sealed tamperproof property bag. There was also no recorded information regarding seizure details or other non-digital forensic investigation (e.g. DNA, fingerprints or ballistic).

Based on our online research, we determined that the DJI Phantom series has a removable memory card within the camera, and a separate memory card fixed to the motherboard. The camera memory card usually stores media (images and videos), the motherboard memory card usually stores flight logs. Our online research also suggested that flight logs can be viewed using open source tools such as DatCon.

We were also not concerned with gathering artefacts to support the elements of proof in this evaluation; hence, the omission of relevant steps.

This study will use EnCase (Version 7.12.01), which has been validated under ISO/IEC 17025.

As part of our examination, we determined that the exhibit has the following characteristics:

1. Device:
   - Model - W322B (DJI Phantom 3)
   - QR code – P*****1 7*****J

2. Battery:
   - Model - PH3-4480mAh-15.2V

3. ID/reference number – 6***********2

4. Video capture facility - No

5. Audio capture facility - No

6. Load carrying capacity – Yes

7. Offensive capability - No

8. Defensive capability - No

9. Is there visible damage to the device? - Yes, one propeller is broken along with minor damage to the propeller arm.

10. Based on the standard fittings (As per the manufactures website), does anything appear to be missing? - Yes, normally there is a camera mounted on the base of the device. However, this appears to have been removed. The mounting point and cables remain, indicating that it was once attached.

11. Exhibit measurements:

    - 380mm by 380mm

12. Identified modifications (Where possible, when compared with factory default options):

    - Non-standard battery - No (As listed on the DJI website)

    - Non-standard motors - No (When visually compared on the DJI website)

    - Non-standard propellers - No (When visually compared on the DJI website)

    - Non-standard camera - Standard camera removed, no camera present.

    - Load carrying device - What appears to be fishing wire, tired between the landing struts, with a large amount remaining. The remaining amount could be used to carry a payload.

13. Identified digital storage (see Figure 3):

    - External memory card - Not present - Normally present but removed with the camera.

    - Internal memory card - 4GB SanDisk Micro SD card.

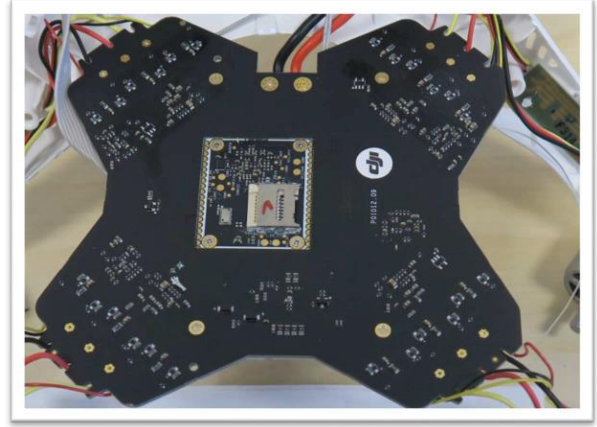

Figure 3. UAV's digital storage

14. Identified Ports:

    - Micro USB

15. Peripheral devices:

    - No other devices submitted.

16. Internal 4GB Micro SD Card removed.

17. E01 created using EnCase - Complete. Hashes match and this also revealed no bad sectors. Carving using EnCase - Complete. EnCase used to view data within .DAT files. Files contain dates but no other legible data (see Figure 4).

| Name | File Ext | Logical Size | Category | File Created |
|---|---|---|---|---|
| FLY095.DAT | DAT | 37,584,896 | Windows | 27/11/16 23:38:04 |
| FLY096.DAT | DAT | 143,818,752 | Windows | 28/11/16 22:33:54 |
| FLY097.DAT | DAT | 177,045,504 | Windows | 29/11/16 01:50:28 |
| FLY098.DAT | DAT | 96,960,512 | Windows | 29/11/16 19:33:28 |
| FLY099.DAT | DAT | 102,957,056 | Windows | 02/12/16 17:57:00 |
| FLY100.DAT | DAT | 209,092,608 | Windows | 02/12/16 22:44:04 |
| FLY101.DAT | DAT | 158,728,192 | Windows | 04/12/16 15:49:34 |
| FLY102.DAT | DAT | 85,819,392 | Windows | 04/12/16 17:57:08 |
| FLY103.DAT | DAT | 41,058,304 | Windows | 04/12/16 23:41:22 |
| FLY104.DAT | DAT | 0 | Windows | 10/12/16 19:59:56 |

Figure 4. .DAT files

18. DatCon 2.4.0 used to interpret flight log data – Complete (see Figure 5).

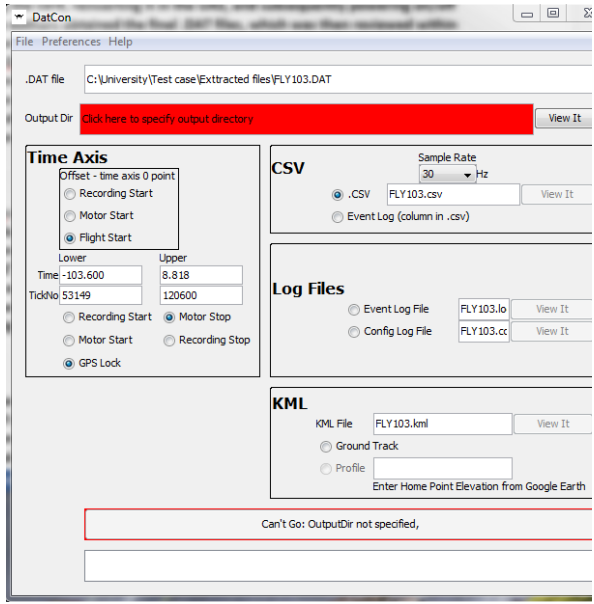

Figure 5. Interpret flight log data

Flight log data obtained, which contained dates, GPS locations, etc.

19. The final .DAT file (FLY104.DAT) contained data, but it could not be viewed using DatCon. Our online research suggested that the file was not closed correctly, and that the DJI Phantom 3 does not close a .DAT file until it is required to open another.

20. EnCase was then used to clone Micro SD Card, and the cloned Micro SD Card was placed in the UAV. The UAV was first turned ON, and then turned OFF.

    The cloned Micro SD card was viewed using EnCase. The final .DAT file was extracted and viewed using DatCon successfully. All relevant data were extracted.

    Micro SD card was placed back into the case exhibit.

21. After using DatCon, two files were created, namely:

    - FLY104.csv contained a spreadsheet (Viewable in Excel) listing all the flight record data. Notably this spreadsheet contained GPS data, battery capacity and height.
    - FLY104.kml, when combined with Google Earth, plots the route of the UAV. When properly set up, one should be presented with an accurate flight path.

22. The summary of data obtained after analysis and its relevance to an investigation are as follows.

    The UAV has minor damage, which includes damage to one of the four propellers. This would indicate that the device suffered damage on impact and would support the assertion that it was in flight immediately prior to the crash.

    The device had been modified by the removal of the camera and the addition of a load carrying mechanism. The removal of the camera also meant that the memory card (located in the camera mount), which normally contains media, was not present.

    The removal of the camera could also be an indication that the user wished to minimize the likelihood of attribution, or reduce weight in order to allow a larger payload.

    The addition of a load carrying mechanism indicates that the device had been adapted specifically to carry a payload. Thus, it would support the assertion that the device was being used to carry a payload prior to its crash.

    The retrieved flight logs contained the final flight, along with previous flight logs. The previous flight logs could contain evidentially useful information, which could include the home address of the user, along with friends and associates. Previous flight logs could also contain historical offences.

## 5. CONCLUSION AND FUTURE WORK

UAVs will play an increasingly important role in future digital (forensic) investigations, as such devices become more sophisticated and their usage become more common in our society.

In this paper, we presented a UAV focussed forensic investigation process, and used it to guide the investigation of the DJI Phantom 3 drone.

Future research will include extending the work in this paper to forensically examine other UAV models and makes, and possibly obtain a taxonomy of forensic artefacts that can be recovered from such devices (similar to the approach of Afzar et al [20][21][22]). We also look at the possibility of adapting the proposed process in the vehicle forensics [27].

As previously discussed, one limitation in UAV forensics is the lack of validated forensically sound tools; hence, this is another potential research direction. For example, the next logical step would be to create some form of parsing tool that could analyse original data and provide a readable and reliable result. Besides, UAV could be integrated with radio communication services in the future. Hence, forensic acquisition and analysis of artefacts from radio-communication services [28] can also be explored.

Finally, anti-UAV forensics is also another potential topic of research interest. We need to understand the types of activities and their effectiveness that may be undertaken by cybercriminals to counter forensic investigations.